\documentclass[twocolumn,showpacs,preprintnumbers,amsmath,amssymb,nofootinbib]{revtex4}

\usepackage{graphicx}
\usepackage{dcolumn}
\usepackage{bm}

\begin{document}

\preprint{CERN-PH-TH/2004-117, CPT-2004/P.041}

\title{Topological susceptibility in the SU(3) gauge theory}

\author{Luigi Del Debbio$^a$, Leonardo Giusti$^b$, Claudio Pica$^c$}

\affiliation{\vspace{0.3cm}
\vspace{0.1cm} $^a$ CERN, Department of Physics, TH Division, CH-1211 Geneva 23, Switzerland\\
$^b$ Centre de Physique Th\'eorique, Case 907, CNRS
Luminy, F-13288 Marseille Cedex 9, France\footnote{
UMR 6207 - Unit\'e Mixte de Recherche du CNRS et des
Universit\'es Aix-Marseille I, Aix-Marseille II
et du Sud Toulon-Var - Laboratoire affili\'e \`a la FRUMAM}\\
\vspace{0.1cm} $^c$ Dipartimento di Fisica dell'Universit\`a di Pisa and 
INFN, Via Buonarroti 2, I-56127 Pisa, Italy}

\date{\vspace{0.2cm} \today}

\begin{abstract}
We compute the topological susceptibility for the SU(3) Yang--Mills
theory by employing the expression of the topological charge density
operator suggested by Neuberger's fermions. In the continuum limit we 
find $r_0^4\chi = 0.059(3)$, which corresponds to $\chi=(191 \pm 5 \,\mathrm{MeV})^4$ 
if $F_K$ is used to set the scale. Our result supports the Witten--Veneziano 
explanation for the large mass of the $\eta'$.   
\end{abstract}

\maketitle

\section{Introduction\label{sec:intro}}
\vspace{-0.2cm}
The topological susceptibility in the pure Yang--Mills (YM) gauge
theory can be formally defined in Euclidean space-time as
\begin{equation}
\chi = \int d^4x\: 
\langle q(x) q(0) \rangle \; ,
\label{eq:chidef}
\end{equation}
where the topological charge density $q(x)$ is given by
\begin{equation}
q(x)=-\frac{1}{32\pi^2} 
\epsilon_{\mu\nu\rho\sigma} \mathrm{Tr}\Big[F_{\mu\nu}(x) F_{\rho\sigma}(x)\Big]\; .
\end{equation}
Besides its interest within the pure gauge theory, $\chi$
plays a crucial r\^ole in the QCD-based explanation of the large 
mass of the $\eta'$ meson proposed  by Witten and Veneziano (WV) a long time ago
\cite{Witten:1979vv,Veneziano:1979ec}. The WV mechanism
predicts that at the leading 
order in $N_\mathrm{f}/N_\mathrm{c}$, where $N_\mathrm{f}$ and $N_\mathrm{c}$ 
are the number of flavors and colors respectively,
the contribution due to the anomaly to the mass of the 
$U_{\mathrm{A}}(1)$ particle is 
given by \cite{Witten:1979vv,Veneziano:1979ec,Seiler:1987ig,Giusti:2001xh,Seiler:2001je} 
\begin{equation}
\displaystyle \frac{F_{\pi}^2 m^2_{\eta^\prime}}{2 N_\mathrm{f}} = \chi\, ,  
\label{eq:WV}
\end{equation}
where $F_{\pi}$ is the corresponding pion decay 
constant\footnote{In our conventions, the physical 
pion decay constant is $92$~MeV.}. Notice that Eq.~(\ref{eq:WV})
is expected to be exactly satisfied if the l.h.s. is computed in 
full QCD and the r.h.s. in the pure gauge theory, both in the 
't~Hooft large-$N_{\mathrm{c}}$ limit~\cite{'tHooft:1973jz}.\\
\indent The lattice formulation of gauge theories is at present 
the only approach where non-perturbative computations 
can be performed with controlled systematic errors. 
Recent theoretical developments~\cite{Kaplan:1992bt,Narayanan:1993wx,Narayanan:1994sk,Furman:1995ky}
(for a recent review see \cite{Giusti:2002rx}) led to the discovery of a fermion 
operator \cite{Neuberger:1997fp,Neuberger:1997bg,Neuberger:1998wv} 
that satisfies the 
Ginsparg--Wilson (GW) relation \cite{Ginsparg:1982bj}, and therefore  
preserves an exact chiral 
symmetry at finite lattice spacing~\cite{Luscher:1998pq} 
\begin{equation}
\psi\rightarrow \widehat \gamma_5\psi\, , \qquad \bar \psi\rightarrow \bar \psi \gamma_5 \, ,
\end{equation}
where $\widehat \gamma_5 = \gamma_5 (1-\bar a
D)$, $D$ is the massless Dirac operator and $\bar a$ is proportional 
to the lattice spacing (see below). The corresponding Jacobian is 
non-trivial~\cite{Luscher:1998pq}, and the chiral anomaly is recovered {\it \`a la} Fujikawa 
\cite{Fujikawa:1979ay} with the topological charge density operator defined 
as\footnote{We use the same notation for analogous quantities in the continuum and 
on the lattice, since they can be clearly distinguished from the context.} \cite{Hasenfratz:1998ri}:
\begin{equation}\label{eq:qx}
q(x) = -\frac{\bar a}{2}\, \mathrm{Tr}\Big[\gamma_5 D(x,x)\Big] , 
\end{equation}
where the trace runs over spin and color indices.
These developments triggered a breakthrough in 
the understanding of the topological properties of the YM 
vacuum. They made it possible to find an unambiguous definition of 
the topological susceptibility with a 
finite continuum limit~\cite{Giusti:2001xh,Giusti:2004qd,Luscher:2004fu},
which is independent of the details 
of the lattice definition~\cite{Luscher:2004fu}. If the charge density 
suggested by GW fermions $Q \equiv \sum_x q(x) = n_+ - n_-$, 
with $n_+$ ($n_-$) the number of 
zero modes of $D$ with positive (negative) chirality in a given background,
is employed, 
the suggestive formula
\begin{equation}\label{eq:chil}
\chi = \lim_{\tiny \begin{array}{c}a\rightarrow 0\\V\rightarrow \infty\end{array}}
       \frac{\langle Q^2 \rangle}{V}
\end{equation}
is recovered, where $V$ is the volume. 
An immediate consequence is an unambiguous derivation of 
the WV formula~\cite{Giusti:2001xh} which, thanks to new simulation 
algorithms~\cite{Giusti:2002sm},
allows for a non-perturbative investigation of the WV mechanism 
with controlled systematics.\\ 
\indent In the past the topological properties of the pure 
gauge theory were investigated with fermionic~\cite{Bochicchio:1984hi,Smit:1987fn} 
and bosonic methods~\cite{Berg:1981nw,Luscher:1982zq,Teper:1985rb,Alles:1997nm,deForcrand:1997ut,Hasenfratz:1998qk,Lucini:2001ej,AliKhan:2001ym,DelDebbio:2002xa}.
These results, however, are affected by model-dependent systematic errors that are not quantifiable, 
and their interpretation rests on a weak theoretical ground. 
Several exploratory computations have already studied the
susceptibility employing the GW definition of the topological 
charge~\cite{Edwards:1998wx,DeGrand:2002gm,Gattringer:2002mr,Cundy:2002hv,Hasenfratz:2002rp,Chiu:2003iw,DelDebbio:2003rn,Giusti:2003gf}.\\ 
\indent The aim of this work is to 
achieve a precise and reliable determination of $\chi$ in 
the continuum limit. In order to reach a robust estimate
of the error on the extrapolated value, we supplement the most recent 
and accurate results~\cite{DelDebbio:2003rn,Giusti:2003gf} with additional simulations,
and we perform a detailed analysis of the various sources of systematic
uncertainties. The result for the
adimensional scaling quantity computed on the lattice is $r_0^4
\,\chi=0.059(3)$, where $r_0$ is a low-energy reference scale
\cite{Guagnelli:1998ud}. In physical units, it corresponds to 
$\chi=(191 \pm 5 \,\mathrm{MeV})^4$ if $F_K$ is used to set the scale.
Our result supports the WV explanation for the large mass of 
the $\eta^\prime$ meson within QCD.

\section{Lattice computation \label{sec:latt}}
\begin{figure}[t]
\begin{center}
\includegraphics*[width=8.0cm,height=7.0cm,angle=0]{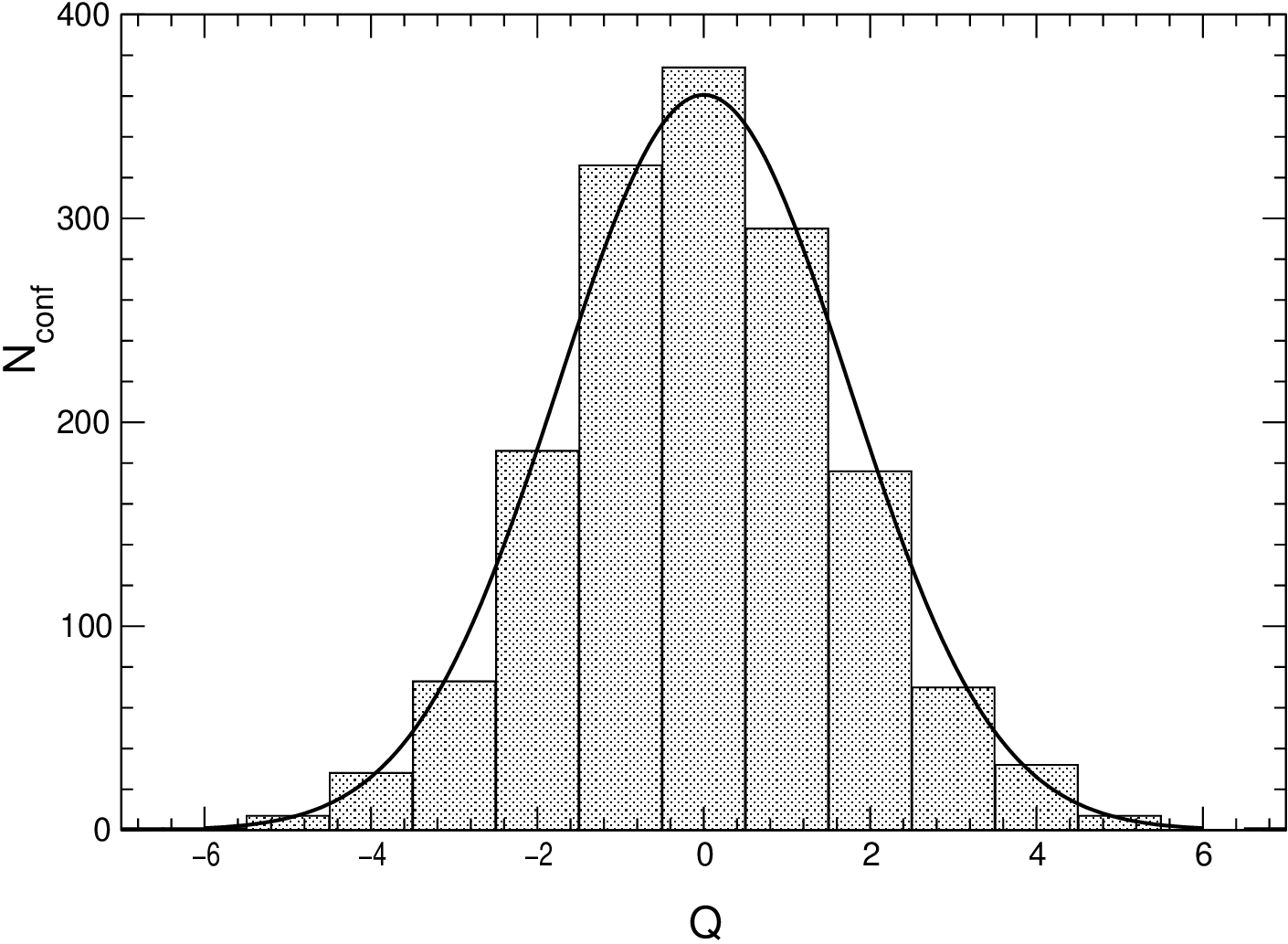}
\caption{\label{fig:histoD} Histogram for the distribution of the
topological charge $Q$ from the lattice $D$.}
\end{center}
\vspace{-0.50cm}
\end{figure}

\begin{figure}
\begin{center}
\includegraphics*[width=8.0cm]{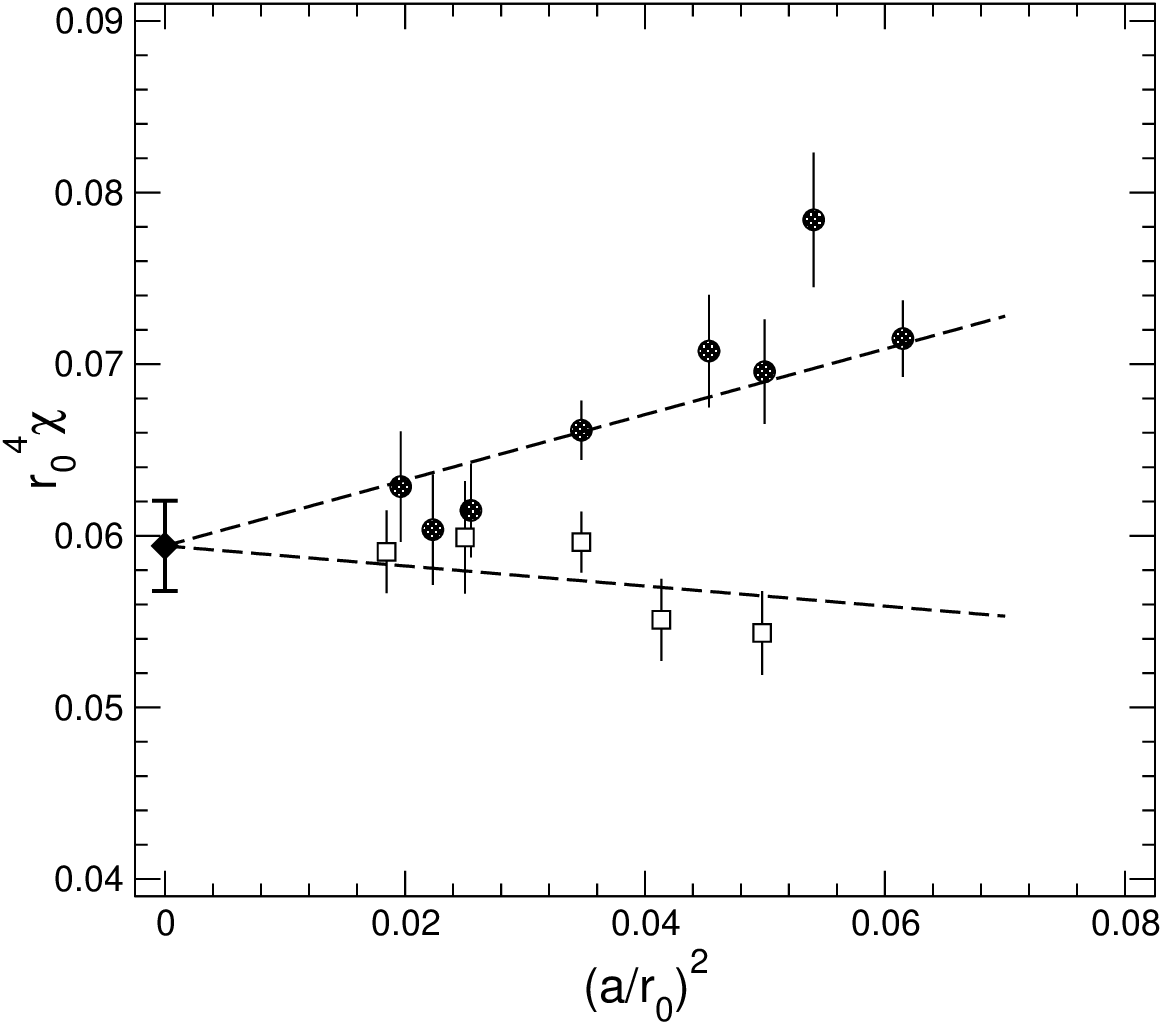}
\caption{\label{fig:continuum} Continuum extrapolation of the adimensional
  product $r_0^4 \chi$. The $s=0.0$ and $s=0.4$ data sets are
  represented by black circles and white squares respectively. 
  The dashed lines represent the results of the
  combined fit described in the text. The filled diamond at $a=0$ is the
  extrapolated value in the continuum limit.}
\end{center}
\vspace{-0.50cm}
\end{figure}

\vspace{-0.2cm}
The numerical computation is performed by standard Monte Carlo
techniques. The ensembles of gauge configurations are generated 
with the standard Wilson action and periodic boundary conditions,
using a combination of heat-bath and over-relaxation updates.
More details on the generation of the gauge configurations can be 
found in Refs.~\cite{DelDebbio:2003rn,Giusti:2003gf}. 
Table~\ref{tab:lattices} shows the list of simulated lattices, where
the bare coupling constant $\beta=6/g_0^2$, the linear size $L/a$ in
each direction and the number of independent configurations are
reported for each lattice.\\
\indent The topological charge density is defined as in 
Eq.~(\ref{eq:qx}), with
$D$ being the massless Neuberger--Dirac operator:
\begin{eqnarray}
D & = & \frac{1}{\bar a} \Big[1 + \gamma_5 \mathrm{sign}(H)\Big]\\
H & = & \gamma_5 (a D_\mathrm{w} -1 -s)\, , \qquad \bar a = \frac{a}{1+s}\, .
\label{eq:overlap}
\end{eqnarray}
Here $s$ is an adjustable parameter in the range $|s|<1$, and
$D_\mathrm{w}$ denotes the standard Wilson--Dirac operator (the
notational conventions not explained here are as in
Ref.~\cite{Giusti:2002sm}). For a given gauge configuration, 
the topological charge is computed by counting the number of 
zero modes of $D$ with the algorithm proposed in Ref.~\cite{Giusti:2002sm}.
As $s$ is varied, $D$ defines a one-parameter family of fermion discretizations, 
which correspond to the same continuum theory but with different
discretization errors at finite lattice spacing. Our analysis includes 
data sets computed for $s=0.4$ and $s=0.0$. Most
of the data were taken from Refs.~\cite{Giusti:2003gf} 
and \cite{DelDebbio:2003rn} for $s=0.4$ and  $s=0.0$ respectively.
The number of configurations were increased, where necessary, in order 
to achieve homogeneous statistical errors of the order of 5\% for each data point. 
Some new lattices were added so as to perform careful studies of the
systematic uncertainties which we describe below, before presenting 
the physical results.\\
\begin{table}[t]
\begin{center}
\setlength{\tabcolsep}{.25pc}
\begin{tabular}{llcccrcc}
\hline
lat    &$\beta$&$L/a$&$r_0/a$&$L$[fm]&$N_{\mathrm{conf}}$& $\langle Q^2\rangle$ & $r_0^4\chi$ \\[0.125cm]
\hline
${\rm A}_1$&$6.0$   &$12$&$5.368$&$1.12$ &$2452$&$1.633(48)$&$0.0654(22)$\\[0.125cm]
${\rm A}_2$&$6.1791$&$16$&$7.136$&$1.12$ &$1138$&$1.589(76)$&$0.0629(32)$\\[0.125cm]
${\rm A}_3$&$5.8989$&$10$&$4.474$ &$1.12$&$1460$&$1.737(72)$&$0.0696(30)$\\[0.125cm]
${\rm A}_4$&$6.0938$&$14$&$6.263$&$1.12$&$1405$ &$1.535(63)$&$0.0615(27)$\\[0.125cm]
${\rm B}_0$&$5.8458$&$12$&$4.032$&$1.49$&$2918$ &$5.61(16)$ &$0.0715(22)$\\[0.125cm]
${\rm B}_1$&$6.0$   &$16$&$5.368$&$1.49$&$1001$ &$5.58(28)$ &$0.0707(37)$\\[0.125cm]
${\rm B}_2$&$6.1366$&$20$&$6.693$&$1.49$&$963$  &$4.81(24)$ &$0.0604(32)$\\[0.125cm]
${\rm B}_3$&$5.9249$&$14$&$4.697$&$1.49$&$1284$ &$5.59(24)$ &$0.0708(33)$\\[0.125cm]
${\rm C}_0$&$5.8784$&$16$&$4.301$&$1.86$&$1109$ &$15.02(72)$&$0.0784(39)$\\[0.125cm]
${\rm C}_1$&$6.0$   &$20$&$5.368$&$1.86$&$931$  &$12.76(95)$&$0.0662(50)$\\[0.125cm]
${\rm D}$  &$6.0$   &$14$&$5.368$&$1.30$&$1577$ &$3.01(12)$ &$0.0651(27)$\\[0.125cm]
\hline
${\rm E}$&$5.9$   &$12$&$4.483$&$1.34$&$1349$&$2.79(12)$ &$0.0543(24)$\\[0.125cm]
${\rm F}$&$5.95$  &$12$&$4.917$&$1.22$&$1291$&$1.955(79)$&$0.0551(24)$\\[0.125cm]
${\rm G}$&$6.0$   &$12$&$5.368$&$1.12$&$3586$&$1.489(37)$&$0.0596(18)$\\[0.125cm]
${\rm H}$&$6.1$   &$16$&$6.324$&$1.26$&$962$ &$2.45(13)$ &$0.0599(33)$\\[0.125cm]
${\rm J}$&$6.2$   &$18$&$7.360$&$1.22$&$1721$&$2.114(76)$&$0.0591(24)$\\[0.125cm]
\hline
\end{tabular}
\caption{
\label{tab:lattices} Simulation parameters and results. For lattices ${\rm A}_1$--${\rm D}$
and ${\rm E}$--${\rm J}$, $s=0.4$ and $s=0.0$ respectively.}
\end{center}
\vspace{-0.75cm}
\end{table}
\indent In order to compute its autocorrelation 
time, we monitor the topological charge determined with the index of $D$ 
for 500 update cycles (1 heat-bath and 6
over-relaxation of all link variables) for the lattice $A_1$.
The autocorrelation time, $\tau_{Q}$, estimated as in Ref.~\cite{DelDebbio:2002xa},
turns out to be compatible with the one obtained for the same lattice by defining the 
topological charge with the cooling technique adopted in Ref.~\cite{DelDebbio:2002xa}.
Based on the experience with cooling, where longer Monte Carlo histories can be analyzed,
we estimate $\tau_{Q}$ for all our lattices; 
for each run we separate subsequent measurements by a
number of update cycles 1--2 orders of magnitude
larger than the estimated $\tau_{Q}$ at the corresponding value of $\beta$.
Statistical errors are thus computed assuming that the measurements are
statistically independent.\\
\begin{figure}[b]
\vspace{-0.25cm}
\begin{center}
\includegraphics*[width=8.0cm]{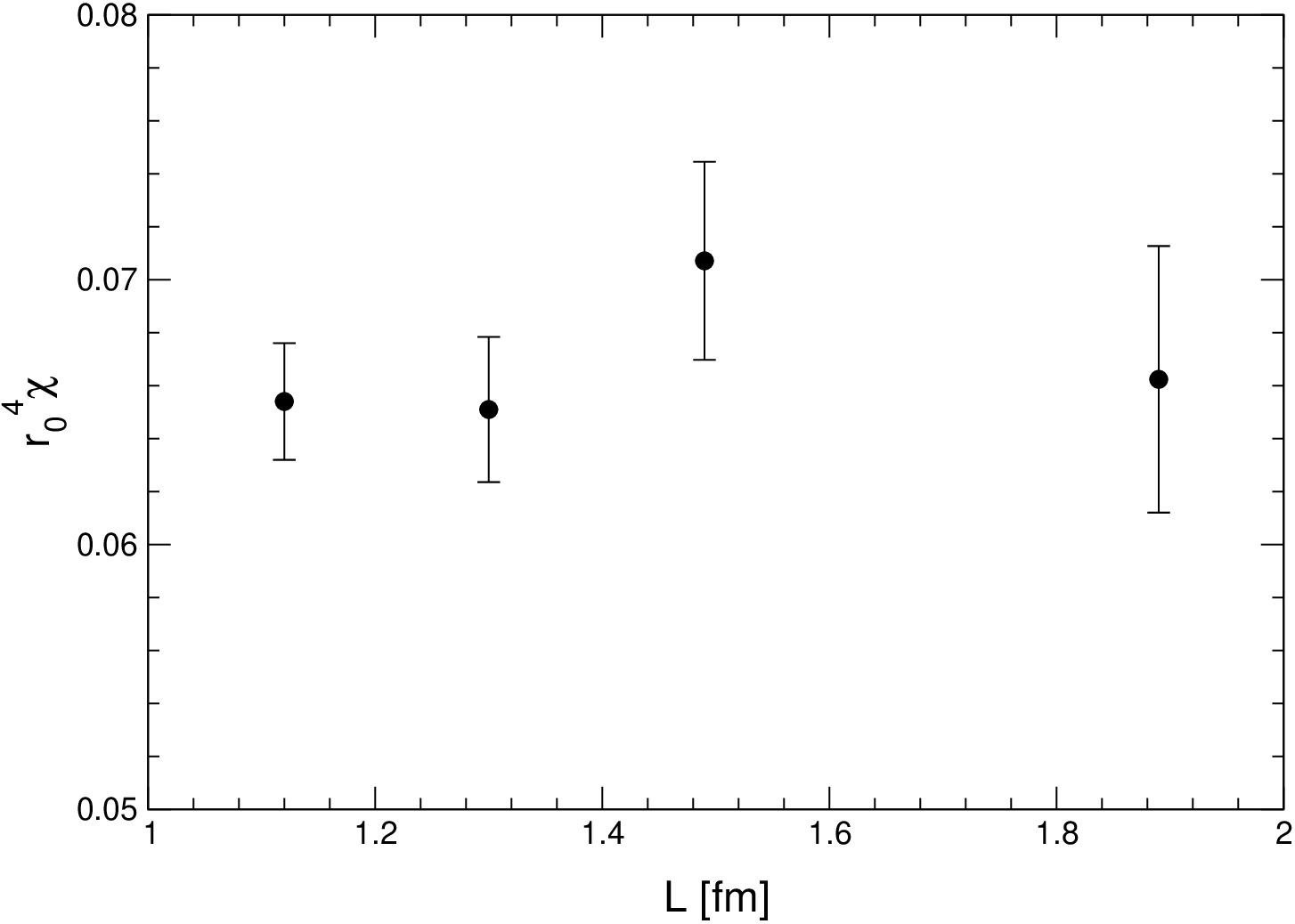}
\caption{\label{fig:volume} The topological susceptibility, in units of
$r_0^{-4}$, as a function of the linear lattice size, in fm, at $\beta=6.0$.}
\end{center}
\end{figure}
\indent  Besides the statistical errors, the
systematic uncertainties stem from finite-volume effects and from the
extrapolation needed to reach the continuum limit.\\
\indent  The pure gauge
theory has a mass gap, and therefore the topological susceptibility
approaches the infinite-volume limit exponentially fast with
$L$. Since the mass of the lightest glueball is around 1.5 GeV, finite-volume 
effects are expected to be far below our statistical errors as
soon as $L\ge 1$~fm. In order to further verify that no sizeable finite-volume
effects are present in our data, we simulated four lattices at
$\beta=6.0$ but with different linear sizes
$L=1.12,1.30,1.49,1.86$~fm. The results obtained for $\chi$
are shown in Fig.~\ref{fig:volume}, where no dependence on $L$
is visible, hence confirming that finite-volume effects are below 
our statistical errors. In the large-volume regime the
probability distribution of the topological charge is expected to be a
Gaussian of the form~\cite{Giusti:2003gf}
\begin{equation}
P_Q = \frac{1}{\sqrt{2\pi \langle Q^2\rangle}} 
e^{-\frac{Q^2}{2\langle Q^2 \rangle}} \; .
\end{equation}
We have checked that this formula describes  
all our data samples very well; for the lattice $D$, the results
are shown in Fig.~\ref{fig:histoD}. Much higher statistics are
required in order to highlight the deviations from a Gaussian
distribution; higher momenta of the topological charge distribution
measured on our data are all compatible with zero within large
statistical errors.\\
\indent As pointed out in the introduction, the topological 
susceptibility defined from the index of the Neuberger operator is 
not plagued by power divergences and does not require multiplicative 
renormalization. This is a distinctive feature
of this approach, which is at variance with what happens for other 
definitions used in the past to compute $\chi$.
At finite lattice spacing, $\chi$ is affected by discretization effects
starting at $O(a^2)$, which are not universal, and, in our case, depend
on the value of $s$ chosen to define the Neuberger operator. 
In order to compare results at different lattice spacings, and to 
extrapolate them to the continuum limit, we 
adopt $r_0$ as the reference scale; this choice is motivated by 
its precise determination in the range of $\beta$ explored 
in this work~\cite{Guagnelli:1998ud}. The values of the adimensional quantity $r_0^4 \chi$ that we obtain
are reported in Table~\ref{tab:lattices}. Data, displayed in Fig.~\ref{fig:continuum} as a 
function of $a^2/r_0^2$, show sizeable 
$O(a^2)$ effects for both the $s=0.4$ and $s=0.0$ samples. 
For $\beta \leq 6.0$, the difference
between the two discretizations is statistically significant.
Within our statistical errors, and in the range where our 
simulations are performed, our results
suggest a linear dependence in $a^2$.  For the $s=0.4$ sample, the value of 
$\chi^2$ per degree of freedom, $\chi^2_{\mathrm{dof}}$,
clearly disfavors a constant behavior, while a linear fit of the form
\begin{equation}
r_0^4 \chi(s) = c_0 + c_1(s) \Big(\frac{a}{r_0}\Big)^2 
\end{equation}
yields a value of $c_0=0.056(3)$ with $\chi^2_{\mathrm{dof}}\approx 0.79$.  
The quadratic fit in  $a^2/r_0^2$ yields an extrapolated value 
compatible with that of the linear one, but with an error three times 
larger, and the coefficient of the quadratic term compatible with zero.
For the $s=0.0$ sample, all three fits give good values of 
$\chi^2_{\mathrm{dof}}$, and for the linear one we obtain
$c_0=0.064(4)$ with  $\chi^2_{\mathrm{dof}}\approx 0.68$, which is compatible with the outcome of the 
same fit for $s=0.4$. The agreement between the two extrapolations 
indicates that we reached the scaling regime. This is confirmed by the 
compatibility of the results in the two data sets for $\beta>6.0$. A robust 
estimate of $\chi$ in the continuum limit can thus 
be obtained by performing a combined linear fit of the data. This fit gives
a very good value of $\chi^2_{\mathrm{dof}}$ when all sets are included, 
and is very stable if some points at larger values of $a^2/r_0^2$ are removed. 
In particular a combined fit of all points with $a^2/r_0^2<0.05$ gives $c_0=0.059(3)$ with $\chi^2_{\mathrm{dof}}\approx 0.73$, and the error is expected to be Gaussian.

\vspace{-0.25cm}
\section{Physical results \label{sec:phys}}
\vspace{-0.2cm}
From the previous analysis, our best result for the topological 
susceptibility is the one obtained from a combined fit of the two sets of
data with $a^2/r_0^2<0.05$:
\begin{equation}\label{eq:finale}
r_0^4 \chi = 0.059 \pm 0.003\, , 
\end{equation}
which is the main result of this work. 
Since $r_0$ is not directly accessible to experiments, we 
express our result in physical units by using the lattice determination
of $r_0 F_K=0.4146(94)$ in the pure gauge theory with valence quarks~\cite{Garden:1999fg} 
and, taking $F_K=160(2)$~MeV as an experimental input, we obtain
\begin{equation}
\chi = (191 \pm 5\, \mathrm{MeV})^4 \, ,
\end{equation}
which has to be compared with \cite{Veneziano:1979ec}
\begin{equation}\label{eq:exp}
\frac{F_\pi^2}{6} \Big(m^2_\eta + m^2_{\eta '} - 2 m^2_K\Big)\Big|_{\mathrm{exp}} \simeq 
(180\, \mathrm{MeV})^4 \; .
\end{equation}
Notice that, since Eq.~(\ref{eq:WV}) is valid only at the 
leading order in a $N_{\mathrm{f}}/N_{\mathrm{c}}$ expansion,
the ambiguity in the conversion to physical units in the pure
gauge theory is of the same order as the neglected terms.\\
\indent Our result supports the fact that the bulk of the mass of the pseudoscalar
singlet meson is generated by the anomaly through the Witten--Veneziano 
mechanism.
\vspace{-0.5cm}
\section*{Acknowledgments\label{sec:ack}}
\vspace{-0.25cm}
It is a pleasure to thank M.~L\"uscher, G.~C.~Rossi, R.~Sommer, M.~Testa, G.~Veneziano and
E.~Vicari 
for interesting discussions. Many thanks also to P.~Hern\'andez, M.~Laine, M.~L\"uscher,
P.~Weisz and H.~Wittig for allowing us to use data on the topological susceptibility
generated in Refs.~\cite{Giusti:2003gf,Giusti:2003iq}. The simulations were
performed on PC clusters at the Cyprus University, 
the Fermi Institute of Rome and at the Pisa University.
We wish to thank all these institutions for supporting our project
and the staff of their computer centers (particularly M. Davini and F. Palombi) for their help.
L.~G.~thanks the CERN Theory Division, where this work was completed, for the warm hospitality 
and acknowledges partial support by the EU under contract HPRN-CT-2002-00311 (EURIDICE).

\bibliographystyle{apsrev}
\bibliography{Q2}
\end{document}